\documentstyle[12pt]{article}
\newcommand{\bm}{\bibitem}
\newcommand{\ud}{\bf}
\begin{document}
\normalbaselineskip = 16 true pt
\normalbaselines
\bibliographystyle{unsrt}

\def\be {\begin{equation}}
\def\ee {\end{equation}}
\def\bea {\begin{eqnarray}}
\def\eea {\end{eqnarray}}

\thispagestyle{empty}
\rightline{\large\sf SINP-TNP/95-08}

\rightline{\large June 1995}

\begin{center}
{\Large\bf{Tensor coupling and  vector mesons\\
in dense nuclear matter}}\\[5mm]
{\sl Abhee Kanti Dutt- Mazumder \footnote{E-mail:
abhee@saha.ernet.in},  Binayak Dutta-Roy \footnote{E-mail:
 bnyk@saha.ernet.in},\\
 Anirban Kundu \footnote{E-mail: akundu@saha.ernet.in}
and Triptesh De}\\[5mm]
Theory Group,\\
Saha Institute of Nuclear Physics, 1/AF Bidhannagar,\\
Calcutta - 700 064, India\\
\end{center}
\medskip
\begin{abstract}

The effects of magnetic interaction between vector mesons and nucleons
on the propagation (mass and width) of the $\rho$-meson in particular
moving through very dense nuclear
 matter is studied and the modifications, qualitative
and quantitative, due to the relevant collective modes (zero-sound and
plasma frequencies) of the medium discussed.
It is shown that the $\rho$-mesons
produced in high-energy nuclear collisions will be
longitudinally polarized in the region of sufficiently dense nuclear matter,
 in the presence of such an interaction.

\end{abstract}
\newpage

\noindent{\bf{1. Introduction}}\\

The study of the properties of hadrons in hot and dense nuclear
matter is of cardinal importance in the understanding of various signals
that probe the dynamics of heavy ion
collisions \cite{kura,ev}, as also in the study of
the equation of state of superdense matter.
Such investigations furthermore are of relevance to various important
issues of nuclear astrophysics dealing with the properties of
neutron stars, cooling of supernovae, the gravitational collapse of
massive stars etc \cite{ruivo,ge}.
 Despite numerous theoretical attempts to
determine the properties of vector mesons in dense nuclear
matter, controversy still exists regarding their in-medium effective
masses and decay widths \cite{je};  various
formalisms such as the
NJL model, Walecka model or QCD sum rules \cite{ho,sh,saito,hat} give
different results \cite{koi,as,br}.
Calculation of mass shifts and decay widths of mesons in
nuclear matter has assumed particular importance in view of proposals
to make such measurements at
CEBAF \cite{cebaf}.
The essential focus of this paper is to investigate the effect of the
magnetic interaction on vector mesons propagating in
dense nuclear matter. In particular we study the case of the
$\rho$ meson where this effect is expected to be more pronounced.

The paper is organised as follows: the formalism is first set forth
and this is succeeded by a discussion of the results.

\bigskip\bigskip
\newpage
\noindent{\bf{2. Formalism}}\\

The vector meson-nucleon interaction Lagrangian may be written as
\begin{equation}
{\cal L}_{int} = g_{\alpha} [{\bar{N}} \gamma _\mu \tau^\alpha
			     N - \frac{\kappa _\alpha}{2M}{\bar{N}}
							 \sigma_{\mu\nu}\tau^\alpha N\partial ^\nu]
V^\mu_\alpha
\end{equation}
where $V_{\alpha} = \{\omega,\rho\}$,  $\alpha$ running
from 0 to 3, indexes quantities relevant for
$\omega$ when $\alpha = 0$, while
$\alpha = $ 1 to 3 refers to the $\rho$ meson; $\tau^0 = 1$ and
$\tau^i$ are the isospin Pauli matrices.  The coupling constants
$g_\rho$, $g_\omega$ and the ``anomalous" or tensor-coupling parameters
$\kappa_\rho$ and $\kappa_\omega$ may be estimated \cite{sh}
 from the Vector Meson Dominance (VMD) of nucleon form-factors or
from the fitting of the nucleon-nucleon interaction data as done
by the Bonn group \cite{bonn}. In view
of the relatively small value of the iso-scalar magnetic moment of
the nucleon as compared to the iso-vector part, the tensor coupling
is more important for the $\rho$ than it is for the $\omega$.
In our calculation the handling of these two mesons runs very
similarly, the only essential difference residing
in the values of the coupling
parameters for the two cases.
{}From the Lagrangian the $VNN$ vertex factor is given by
\begin{equation}
\Gamma^\alpha_\mu = g_{\alpha}[\gamma_\mu\tau^\alpha
-\frac{\kappa _\alpha}{2M}\sigma_{\mu\nu}
\partial^\nu\tau^\alpha]\\
\end{equation}

We shall consider the propagation of vector mesons in dense nuclear
matter at zero temperature. As such we shall be following the usual
methods of relativistic quantum field theory, with the vacuum
replaced by the ground state of nuclear matter at zero temperature,
specified by the Fermi momentum ($k_F$) corresponding to its density. Such
a field theoretic approach to the study of many body problems was
developed, in the context with which we are concerned,
by Matsubara, Galitskii and Migdal \cite{mat,gal,mig},
and the relativistic
generalization provided by Fradkin \cite{frad},
and most importantly in the approach being rather closely followed
in this paper, by Chin \cite{ch}.
Furthermore, the Hartree or Mean-Field energy density for
dense nuclear matter goes as $k_F^6$, while the correlation energy varies
as $k_F^4$ and this leads us to the position that we may use the
mean-field description for nuclear matter at very high densities
(that being our present concern).

The second order polarization tensor $\Pi_{\mu\nu}$,  for the vector meson
arising from the nucleon loop (Fig. 1), is thus calculable from the
Lagrangian and yields
\begin{equation}
\Pi_{\mu\nu}^{\alpha\beta}=\frac{-i}{(2\pi)^4}
\int {d^4}k~{\rm Tr}[i\Gamma^\alpha_\mu iG(k+q)
{i\bar\Gamma^\beta_\nu} iG(k)]
\end{equation}
where $(\alpha,\beta)$ are the isospin indices and $G(k)$ is the
in-medium nucleon propagator
\begin{equation}
G(k) = G_F(k)  + G_D(k)
\end{equation}
where
\begin{equation}
G_F(k) =
(k_\mu \gamma^\mu +M^\ast)[\frac{1}{k^2-M^{\ast2}+i\epsilon}]
\end{equation}
and
\begin{equation}
G_D(k) = (k_\mu \gamma^\mu + M^\ast) [\frac{i\pi}{E^{\ast}(k)}
\nonumber\\ \delta (k_0-E^\ast (k))\theta(k_F-|\vec k |)]
\end{equation}
with $M^\ast$ denoting the effective mass of the nucleon in the
medium and $E^\ast (\vert k \vert) = \sqrt {\vert \vec k \vert^2 +
M^{\ast 2}}$.
The first term in $G(k)$ namely $G_F(k)$
is the same as the free propagator of a spin $\frac{1}{2}$
Fermion, except for the fact that the effective mass of the nucleon
is to be used, while the second part, $G_D(k)$,
involving $\theta(k_F-|\vec k|)$,
arises from Pauli blocking, describes the modifications
of the same in the nuclear matter at zero temperature \cite{se}, as it
deletes the on mass-shell propagation of the nucleon in nuclear
matter with momenta below the Fermi momentum.

In a similar vein the polarization insertions can also be written as sum of two
parts:
\begin{eqnarray}
\Pi_{\mu\nu}(q)& =& \Pi_{\mu\nu}^F(q) + \Pi_{\mu\nu}^D(q),\\
\Pi_{\mu\nu}^F(q)& =&\frac{-i}{(2\pi)^4}
\int {d^4}k~{\rm Tr}[\Gamma_\mu G_F(k+q)
{\bar\Gamma_\nu} G_F(k)]
\end{eqnarray}

\noindent $\Pi^D_{\mu\nu}(q)$ denotes the density dependent part of the
polarization and $\Pi^F_{\mu\nu}(q)$ denotes the free part i.e. it contains
the effect of `Dirac sea'. The density dependent part has a natural cut
off because of the $\theta(k_F-| \vec k|)$ function whereas the
free part is regularized by adding appropriate counter-terms to the
Lagrangian. As known from ordinary quantum field theory  the
effect of the finite part of $\Pi_F(q)$ is to render the coupling
constant momentum dependent and this modifies the interaction
subtantially as short distances. Here the procedure of regularization
and renormalization is very similar
except that the nucleon mass $M$ would have to be replaced by the
effective nucleon mass in nuclear matter ($M^*$); the tensor coupling,
however, belongs to the genre of unrenormalizable theories and has
to be tackled in the spirit of an effective theory.
In the case of ordinary nuclear matter densities this
short distance modification is important as discussed
 in Ref \cite{sh}, however, in the present
context where we concentrate on the long wavelength limit, i.e. low
momentum transfer to study the collective modes in dense nuclear matter, the
effect of vacuum polarization would in general be small as pointed out
by Chin \cite{ch}.

The real part of the density dependent piece of the polarization is given by
\begin{eqnarray}
\Pi^D_{\mu\nu}
= \frac{g_v^2\pi}{(2\pi)^4}\int\frac{{d^4}k}{E^{\ast}(k)}
\delta (k^0-E^{\ast}(k))\theta
(k_F-\mid \vec k\mid)
\nonumber\\
\cdot \Big[\frac{{\it {T}}_{\mu\nu}(k-q,k)}{(k-q)^2-M^{\ast 2}}
+\frac{{\it {T}}_{\mu\nu}(k,k+q)}{(k+q)^2-M^{\ast 2}}\Big]
\end{eqnarray}

\medskip

Although the form of $\Pi_{\mu\nu}^D(q)$ is the same as that of Chin
\cite{ch}, here, however,
it has three part corresponding to vector-vector, vector-tensor and
tensor-tensor terms.  Hence the self energy can be written as
\begin{equation}
\Pi_{\mu\nu}^D(q) = \Pi_{\mu\nu}^{vv}(q) + \Pi_{\mu\nu}^{vt+tv}(q) +
\Pi_{\mu\nu}^{tt}(q)
\end{equation}
The $\Pi_{\mu\nu}^D(q)$ functions in this case are as follows
\begin{equation}
\Pi_{\mu\nu}^{vv}=\frac{g_v^2}{\pi ^3}\int_0^{k_F}\frac{{d^3k}}{E^{\ast}(k)}
\frac{{\cal K}_{\mu\nu}-Q_{\mu\nu}(k\cdot q)^2}{q^4-4(k\cdot q)^2}
\end{equation}

\begin{equation}
\Pi_{\mu\nu}^{vt+tv}=\frac{g_v^2}{\pi ^3}
(\frac{kM^\ast}{4M})2q^4Q_{\mu\nu}\int_0^{k_F}\frac{{d^3k}}{E^{\ast}(k)}
\frac{{1}}{q^4-4(k\cdot q)^2}
\end{equation}

\begin{equation}
\Pi_{\mu\nu}^{tt}=-\frac{g_v^2}{\pi ^3}
(\frac{k}{4M})^2(4q^4)\int_0^{k_F}\frac{{d^3k}}{E^{\ast}(k)}
\frac{{\cal K}_{\mu\nu}+Q_{\mu\nu}M^{\ast 2}}{q^4-4(k\cdot q)^2}
\end{equation}
\vskip 1 true cm
\noindent where ${\cal K}_{\mu\nu}=(k_\mu-\frac{k.q}{q^2}q_\mu)
(k_\nu-\frac{k.q}{q^2}q_\nu)$ and $Q_{\mu\nu}=(-g_{\mu\nu} +
\frac{q_{\mu}q_{\nu}}{q^2})$.
It is clear that the form for the polarization tensor conforms to the
requirement of current conservation,
i.e.
\be
q_\mu\Pi^D_{\mu\nu}=\Pi^D_{\mu\nu}q_\nu=0
\ee
In order to evaluate $\Pi^D_{\mu\nu}$ conveniently, we choose $\vec k$ to be
along the $x$ axis i.e. $ q=(q_0,\mid\vec q\mid ,0,0) $, and $k.q=\mid k\mid
\mid q \mid \chi -E^{\ast}(k)q_0$, where $\chi$ is the cosine of the
angle between $\vec k$ and $\vec q$. After $\phi$ integration the
non-vanishing components $\Pi^D_{\mu\nu}$ are as shown below

\be
\pmatrix { \Pi_{00} & \Pi_{01} & 0 & 0 \cr
\Pi_{10} & \Pi_{11} & 0 & 0 \cr
0 &  0 & \Pi_{22} & 0 \cr
0 &  0 & 0 & \Pi_{33} }
\ee
\vskip 0.5 true cm

Also for isotropic nuclear matter we have
$\Pi_{22}^D=\Pi^D_{33}$
and $\Pi^D_{01}=\Pi^D_{10}$, and
hence taking all this into account we have only two non-vanishing
independent component of $\Pi^D_{\mu\nu}$, linear combinations of
which gives us the longitudinal and transverse components of
$\Pi^D_{\mu\nu}$, namely,
$\Pi^D_L(q)=-\Pi_{00}^D+\Pi_{11}^D(q)$
 and
$\Pi_T^D(q)=\Pi^D_{22}=\Pi^D_ {33}$.

For collective excitations, the wavelength of the
oscillations must be greater than the interparticle spacing. Thus for
super-dense matter $q<<k_F$ and we can simplify the denominators of
the above integrations by neglecting $q^4$ compared to
$4(k.q)^4$\cite{ch}. In this approximation the integrations can easily be
performed to give results in a closed form, viz.,
\begin{eqnarray}
\Pi_L^D(q_0,{\bf q})
&=&-\frac{g_v^2}{\pi ^2}k_F\epsilon_F\beta(1-c_0^2)\Phi(c_0/v_F)\nonumber\\
\Pi_T^D(q_0,{\bf q})
&=&\frac{1}{2}\frac{g_v^2}{\pi ^2}\frac{k_F^3}{\epsilon_F}
[1 + \beta(1-\frac{c_0^2}{v_F^2})\Phi(c_0/v_F)]\nonumber\\
\beta&=&1-(\frac{\kappa}{2M})^2q^2
\end{eqnarray}
where
\be
 \Phi (x) = -1 + \frac{1}{2}x~\ln\vert \frac{(x+1)}{(x-1)}\vert
\ee
and $c_0 = \frac{q_0}{\vert {\vec q}\vert}$, $v_F
= \frac{k_F}{\epsilon_F}$ and $\epsilon_F$ is the Fermi energy
$\epsilon_F =\sqrt{M^{\ast 2} +k_F^2}$.
In the limit $ \kappa\rightarrow 0 $, of course, the results of Ref\cite{ch}
 ensue.

The effective mass of the $\rho$ mesons in hadronic matter is
described by the poles of the in-medium $\rho$ meson propagator.
For short wavelengths the oscillation of meson fields can be regarded as
the usual meson propagation, as in that regime the effect of Pauli
blocking is not appreciable, but in the long wavelength limit medium effects
can
be substantial and these can be beautifully interpreted
as arising from the meson
`picking up' the collective modes of the nucleonic Fermi fluid \cite{pines}.
 The vector meson propagator is calculated by summing over
ring diagrams, a diagrammatic equivalent of the random phase
approximation (RPA), which consist of repeated insertions of the
lowest order polarization, as illustrated in Fig. 2. We
make use of Dyson's equation to carry out the summation
\be
D_{\mu\nu}(q) = D^0_{\mu\nu}(q) +
D^0_{\mu\alpha}(q)\Pi^{\alpha\beta}(q) D_{\beta\nu}(q)
\ee
The poles are determinable from the equation
\be
det[\delta^\nu_\mu - D^0_{\mu\alpha}\Pi_{\alpha\nu}]=0
\ee
The bracketed term is nothing but the dielectric tensor of the system
\be
\epsilon^\nu _\mu = \delta_\mu^\nu - D^0_{\mu\alpha}\Pi^{\alpha\nu}
\ee
the determinant of which is the dielectric
function of the system . The eigen-condition can now be expressed as
\be
\epsilon (q) = 0 .\ee
\noindent As $\Pi_{\mu\nu}(q)$  is already known,
$\epsilon (q)$ can be calculated immediately . The relevance of the
set of ring diagrams and the origin of such an eigen-condition can be
understood from linear response theory \cite{fet},
where the fluctuation of
the current density, the source term for the meson field in nuclear matter,
given in terms of the polarisation tensor,
is `picked up' by the vector field.

We have already shown that $\Pi_{\mu\nu}^D (q)$ manifestly satisfies the
current conservation conditions, and with the choice of $q_\mu$ already
discussed, assumes a
particularly simple structure eq. (15).  The longitudinal and tranverse
dielectric functions are defined as
\be
\epsilon_T(q) = 1- D^0\Pi^D_{22}=1-D^0\Pi_{33}=1-D^0\Pi^D_T
\ee
\be
\epsilon_L(q) = (1- D^0\Pi^D_{00})(1-D^0\Pi_{11}) -
D^0\Pi^D_{01}D^0\Pi^D_{10}
\ee
The explicit form of the longitudinal and transverse dielectric
functions can be obtained with the help of eqs. (16) and (17). to yield,
\be
\epsilon _T=1+
\frac{1}{q^2-m^2}\frac{1}{2}\frac{g_v^2}{\pi^2}\frac{k_F^3}{\epsilon_F
}[1 + \beta(1-(\frac{c_0}{v_F})^2)\Phi(c_0/v_F)]
\ee
\be
\epsilon _L=1+
\frac{1}{q^2-m^2}\frac{g_v^2}{\pi^2}k_F^3\epsilon_F
\beta(1-(\frac{c_0}{v_F})^2)\Phi(c_0/v_F)
\ee
The eigenmodes of the vector meson are given by
\be
\epsilon(q) = \epsilon_T^2(q) \epsilon_L(q)
\ee
corresponding to the three degrees of fredom of a massive vector particle.
The two identical (or degenerate) transverse collective modes are each
given by
\be
\epsilon_T(q)=0
\ee
and the single longitudinal mode by
\be
\epsilon_L(q) = 0
\ee
which yield the relevant dispersion curves. Solutions to be sought may
be classified as space-like ($\vert \vec q \vert > q_0$) or
time-like ($\vert \vec q \vert < q_0$), the latter being the
`particle' modes, while the former corresponds, as we shall see, to zero sound.

 \bigskip\bigskip

\noindent{\bf {3. Results and discussions}}\\

As already stated earlier, for collective excitations the wavelengths
of the collective oscillations must be greater than the interparticle
spacings, to wit, $\vert {\vec q}\vert <<k_F$.
Furthermore, the stability of the collective modes
 requires on the one hand that
the dispersion curves be such that $q_0^2$ be non-negative or else
$q_0$ would be imaginary and accordingly correspond to physically
unacceptable exponentially growing fluctuations; on the other hand
these modes must furthermore be such that they are not dissipated
through nucleon-antinucleon decay (anticipating that the mass of
the rho-meson, for instance, in dense nuclear matter would make this
energetically possible) for the time-like solutions, or the creation
of a nucleon-nucleon-hole pair (that is the absorption of the rho
to lift a nucleon from below the Fermi sea to one above above it,
creating thereby a hole-particle pair) in the case of a space-like
mode.
Accordingly, as has already been discussed in Ref \cite{ch,se}, for the
collective  modes to be undamped the following condition must be fulfilled
$\frac{q_0}{\vert \vec q\vert}
>v_F$ thus we can expand $\Phi $ for $c_0/v_F>>1$ . In this case we
can retain only first few terms of the expansion which gives
$\Phi (c_0/v_F) = \frac{1}{3c_0^2}$ (as $v_F\rightarrow 0 $).
In this limit for the longitudinal modes one arrives at the dispersion
relation
\be
q_0^4(1+\frac{\Omega^2\kappa^2}{4M^2}) -q_0^2(\Omega^2+\vert \vec q
\vert^2 +m^2 + \frac{\Omega^2\kappa^2}{2M^2}\vert \vec q\vert ^2)
+\Omega^2\vert\vec q\vert^2(1+\frac{\kappa^2|\vec q|^2}{4M^2})=0
\ee
from the equation $\epsilon_L = 0$, while for the transverse modes
one obtains
\be
q_0^4(1-\frac{\Omega^2\kappa^2}{8M^2}) -q_0^2(\Omega^2+\vert \vec q
\vert^2 +m^2 - \frac{\Omega^2\kappa^2}{4M^2}\vert \vec q\vert ^2)
-{1\over 2}\Omega^2\vert\vec q\vert^2(1+\frac{\kappa^2|\vec q|^2}{4M^2})=0.
\ee
Here $\Omega$, given by $\Omega^2 = {1\over 3}{ g^2_V\over \pi^2 }
{ k^3_F\over \epsilon_F}$,
can be interpreted as the relativistic
generalisation of the familiar plasma frequency
\cite{ch} with $e^2$ replaced by $g^2_V / 4\pi $.

One of the solutions of the dispersion relations for the longitudinal
mode is such that as $|\vec q|\rightarrow 0$, $q_0\rightarrow 0$.
For such low lying excitations, we have
\be
q_0^2={\Omega^2\vert\vec q\vert^2\over \Omega ^2 + m^2 +
(1+\frac{\Omega ^2\kappa ^2}{2M^2})\vert \vec q\vert^2}
\ee
when $\vert \vec q\vert \rightarrow 0$, $c_0=\frac{q_0}{\vert\vec q\vert}$
approaches a constant. This form of the dispersion relation is
characteristic of acoustic (sound) propagation. At zero temperature
ordinary sound propagation is not
possible in a Fermi fluid. Following Landau,
one therefore identifies this branch of
collective modes as zero sound branch which differ in nature from
ordinary sound in that what is involved is not a breathing of the
Fermi sphere, but its shape change without alteration of the volume,
the limiting Fermi surface having the form of a surface of
revolution elongated in the forward direction of the propagation of
the wave \cite{pines}.

On the other hand branches corresponding to
high lying excitations are obtained for both the longitudinal and
the transverse modes.
In the static limit, these yield
\be
q_{0T}^2 = \frac {m^2+\Omega^2}{1-\frac{\Omega ^2\kappa ^2}{8M^2}}\ee
\be
q_{0L}^2 = \frac {m^2+\Omega^2}{1+\frac{\Omega ^2\kappa ^2}{4M^2}}.
\ee
These modes correspond to collective oscillations `picked up' by the meson
and the branches (transverse and longitudinal) are known as the particle
or mesonic branches.
Here again $\Omega$ is the plasma frequency, but unlike the case discussed
by Chin \cite{ch} where in the static limit $q^2_{\rm 0T} = q^2_{\rm 0L} = m^2
+ \Omega^2$,
the two modes are split.
Though similar in form, our results are different from those of Chin
\cite{ch} because of the inclusion of tensor coupling.

The main difference resides in the fact that in the presence of
the `magnetic' coupling the transverse mode of the mesonic branches
not only lies higher (as is also the case in Chin's discussion) but
is split from the longitudinal branch even in the static limit ($
\vert \vec q\vert = 0$). This is because the interaction between two
`magnetic dipoles' does not possess azimuthal symmetry. Furthermore
the longitudinal branch is substantially lowered and the transverse
mode elevated in high density nuclear
matter compared to the pure vector-coupling case.  The
non-monotonic nature of the dispersion curve for the transverse mode
is an interesting feature which appears in the presence of the magnetic
coupling.
Nevertheless some of these amusing aspects may not be of
much physical relevance for two reasons, namely, for large values of
$q_0$ and $|\vec q|$ the collective mode ceases to have any physical
meaning (the wavelength is less than the interparticle spacing), and
for values of $q_0$ and $|\vec q|$ not so large (but larger than
$2k_F$) the mode is unstable, and the meson decays into a
nucleon-antinucleon pair. For high nuclear matter density, the
transverse mode, even in the static limit, is unstable, as is evident
from Fig. 3. This shows that in the presence of magnetic
coupling, in the regime of high density nuclear matter,
the produced $\rho$ will be entirely longitudinally
polarized.
On the other hand, the
observed flattening of the zero-sound branch indicates a
reduction of the repulsive force between nucleons due to the
magnetic interaction, as is evident from
eq. (31). Of course, the whole analysis has been done for symmetric nuclear
matter; the case where the nuclear matter is asymmetric is under
study.

In summary, one can say that in the presence of the magnetic interaction,
leads, apart from other effects, to a suppression and even the absence
of the transverse mode of $\rho$ in nuclear matter which is dense
enough and this feature should be
observable in high-energy nuclear collisions.

\bigskip
\centerline {\bf Acknowledgements}
\bigskip

The authors thank Professors Ashoke Chatterjee,
Raj Kumar Moitra and Radhey Shyam for useful discussions.

\newpage

\begin{center}
{\bf Figure Captions}
\end{center}

\begin{enumerate}
\item Self-energy diagram of the $\rho$-meson.

\item Diagrammatical representation of the Dyson equation.

\item Different collective modes with and without the magnetic
interaction. The solid lines represent the dispersion curves with
magnetic coupling $\kappa$ taken to be 6, while the dotted curves
correspond to the case without magnetic coupling. The density of
nuclear matter used for illustration is that corresponding to
$k_F=3.2$ fm$^{-1}$. The value of $g_V$
taken is  2.6  and the units used are MeV.

\end{enumerate}

\end{document}